# Title: Ultrafast Kapitza-Dirac effect

**Authors:** Kang Lin[1,2]*, Sebastian Eckart[2], Hao Liang[3], Alexander Hartung[2], Sina Jacob[2], Qinying Ji[2], Lothar Ph. H. Schmidt[2], Markus S. Schöffler[2], Till Jahnke[4], Maksim Kunitski[2], Reinhard Dörner[2]*

**Affiliations:**

[1]School of Physics, Zhejiang Key Laboratory of Micro-Nano Quantum Chips and Quantum Control, Zhejiang University; Hangzhou, 310027, China.

[2]Institut für Kernphysik, Goethe-Universität; Frankfurt am Main, 60438, Germany.

[3]Max Planck Institute for the Physics of Complex Systems; Dresden, 01187, Germany.

[4]Max-Planck-Institut für Kernphysik, Heidelberg, 69117, Germany.

*Corresponding author. Email: lin@atom.uni-frankfurt.de; doerner@atom.uni-frankfurt.de

**Abstract:** Similar to the optical diffraction of light passing through a material grating, the Kapitza-Dirac effect occurs when an electron is diffracted by a standing light wave. In its original description the effect is time-independent. In the present work, we extend the Kapitza-Dirac concept to the time domain. By tracking the spatiotemporal evolution of a pulsed electron wave packet diffracted by a femtosecond ($10^{-15}$ second) standing wave pulse in a pump-probe scheme, we observe so far unseen time-dependent diffraction patterns. The fringe spacing in the observed pattern differs from that generated by the conventional Kapitza-Dirac effect, moreover it decreases as the pump-probe delay time increases. By exploiting this time-resolved diffraction scheme, we gather access to the time evolution of the previously inaccessible phase properties of a free electron.

**One-Sentence Summary:** Harnessing the time-resolved Kapitza-Dirac effect allows for the retrieval of the phase of a free electron wave packet.

**Main Text:** In 1933, few years after Davisson and Germer demonstrated the wave nature of electrons by diffracting an electron beam at a periodic crystal structure (*1*), Kapitza and Dirac suggested that an electron beam can also be diffracted by an immaterial standing light wave (*2*). Nowadays this is referred to as the Kapitza-Dirac effect (*3*). It was shown that this diffraction occurs due to the stimulated Compton back-scattering of photons in the standing wave. During this process, two times the photon momentum $\hbar \boldsymbol{k}_\gamma$ is imparted onto the electrons causing the deflection of the electron beam to discrete angles. In matter optics the Kapitza-Dirac effect is utilized for manipulation of matter waves. For instance, a standing light wave can serve as a coherent beam splitter for particle bunches, including electrons, ions, atoms and molecules (*3–6*). Such splitter is an essential part of particle interferometers (*7*) that nowadays have many applications ranging from extremely sensitive studies of gravitational effects (*8*) to generation of spin-polarized electrons (*9*). Despite of a huge body theoretical work, the conventional Kapitza-Dirac effect on electrons was not observed until 2001 (*10*) due to experimental challenges mainly related to overall weakness of the effect. In this experiment, Freimund *et al.* used a narrowly collimated electron beam, which was produced by an electron gun. The electrons travelled through a standing light wave, which was formed by two counter-propagating laser beams. In the final electron momentum spectrum, Freimund *et al.* observed (*10, 11*) discrete diffraction peaks spaced by two-photon momenta, $2\hbar \boldsymbol{k}_\gamma$, as sketched in Fig. 1A.

The advent of the ultrafast laser technology finally provided the opportunity to address the dynamics of the Kapitza-Dirac photon scattering in an ultrafast pump-probe experiment and explore the time-dependence of the electron diffraction. In addition, strong laser fields can be used for production of the electrons through ionizing neutral atoms or molecules, which are then subject to the Kapitza-Dirac effect in the experiment. Electrons released by such laser fields from atoms or molecules have unique properties which are different from those produced by an electron gun. The former can be considered as a coherent electron wave packet with a broad distribution in momentum space while the latter can be approximated as a plane wave with a momentum-space distribution close to a delta function. The early attempts to employ ultrashort strong laser pulses in studies on the Kapitza-Dirac effect (referred to as "strong-field Kapitza-Dirac effect") dates back to the end of 80s (*12*). In that work, two counter-propagating, 100 picosecond (1 ps = $10^{-12}$ s) laser pulses were used to form a standing light wave. The pulsed standing wave ionized individual atoms in the gas phase, setting the electrons free which were subsequently diffracted by the same standing wave. As a result of a very high photon-scattering rate in the intense light field, the measured electron angular emission distribution consisted of two peaks with a spacing that corresponds to thousands of photon momenta. No peaks spaced by two-photon momenta were observed (see Ref. (*13*) for theoretical modelling predicting such interference fringes). In 2014, a 55 femtosecond standing wave was used to scatter neutral atoms in another experiment (*14*). However, neither of these experiments observed interference fringes, nor did they have access to the temporal properties of the diffraction process.

Here, we report on a fully time-resolved experiment on the Kapitza-Dirac effect, which in the end reveals the phase evolution of an electron wave packet diffracted by a femtosecond standing wave. The concept of our experiment is illustrated in Fig. 1B. In its actual implementation, an electron wave packet is released from a xenon atom upon strong-field ionization using a highly intense, pulsed standing light wave (pump pulse). After a variable time delay, a weaker, non-ionizing, femtosecond standing light wave (probe pulse) is applied to diffract the emitted electron wave packet. The pump and the probe pulses are generated by a commercial Ti:Sapphire laser system (Coherent Legend Elite Duo, 25 fs, 800 nm, 10 kHz) by focusing two counter-propagating laser pulses into the same target volume (*15*). Using a standing wave as the pump locks the birth position of the electron wave packet relative to the field maxima of the probe standing wave (see Supplementary Materials). The three-dimensional momenta of the electrons were measured using a COLTRIMS reaction microscope (*16*) (see Supplementary Materials).

With this experimental arrangement we observe time-dependent interference fringes in the electron momentum distribution. The fringe spacing decreases continuously by increasing the time delay between the pump and probe pulses. This observation is remarkably different from that of the conventional Kapitza-Dirac effect, where the diffraction pattern shows a constant spacing of two-photon momenta. The corresponding results are shown in Fig. 2, which depicts the measured two-dimensional (2D) electron momentum distribution along the laser polarization (z-axis) and propagation directions (x-axis) for various time delays. For the sake of visibility, a gating of |$p_y$| > 0.1 a.u. (atomic units are used unless stated otherwise) is applied to the dataset in order to remove the Coulomb-focusing-induced distortions around the center of the 3-dimensional momentum distribution (*17*, *18*). As a reference, Figure 2A shows the momentum distribution recorded at negative time delay of -1 ps. In this case the probe pulse arrives before the pump pulse such that the electron wave packet, which is released by the pump pulse, does not experience diffraction by the standing wave. The 2D momentum distribution extends along the polarization direction ($p_z$) of the pump pulse resulting from the acceleration of electrons by the pump pulse's electric field. Along the light-propagation direction ($p_x$) the electron momentum distribution is much narrower since it is determined solely by the initial velocity distribution of the electrons upon tunneling. The distribution is very similar to the extensively studied momentum distribution of electrons released by tunnel ionization using a single travelling femtosecond laser pulse (*19*). The Kapitza-Dirac effect manifests itself in the periodic vertical fringes, i.e. an intensity modulation along the light-propagation direction (horizontal axes in Fig. 2). At negative pump-probe delay (Fig. 2A), we find along this direction a close to Gaussian distribution centered at zero momentum with a full width at half-maximum (FWHM) of ~0.25 a.u. At positive time delays (Figs. 2B-D), the electron momentum distribution turns into fringe patterns in the light-propagation direction, and the spacing between diffraction fringes decreases with increasing time delay, as shown in Figs. 2F-H. Specifically, the spacing between the diffraction peaks at 1, 2 and 3 ps is ~0.18, 0.09 and 0.06 a.u., respectively. To find a quantitative relation between the diffraction spacing and the time delay, we performed a continuous scan of the pump-probe delay from 2 to 10 ps. Figure 3A clearly shows that the spacing between the diffraction fringes decreases continuously with increasing time delay. We find that the spacing between the diffraction peaks equals one half the wavelength of the standing light wave $\lambda_{sw}$ time the electron mass $m_e$ (with $m_e$=1 in atomic units) divided by the time delay, i.e.

$$\Delta p_x = m_e \frac{\lambda_{sw}}{2t} \qquad (1)$$

This is surprising since the conventional Kapitza-Dirac effect is time-independent. For the remainder of the paper we term this newly observed phenomenon "time-resolved Kapitza-Dirac effect".

To understand the underlying physics of the time-resolved Kapitza-Dirac effect, we numerically propagated a one-dimensional electron wave packet followed by interaction with the pulsed standing wave (see Supplementary Materials for details). We start with an initial electron wave packet that is defined by the distribution observed for the negative time delay (Fig. 2E) and has a width of about 0.25 a.u. in momentum space. This is a very broad distribution compared to the transverse momentum distribution of a collimated electron beam (which is often even approximated by a delta function in momentum space). Specifically, in Ref. (*10*), the electron beam has a momentum distribution with a width of 2.2×10⁻⁴ a.u. along the light-propagation direction. Subsequently, the initial electron wave packet evolves under field-free condition. At different time delays, we apply an ultrashort spatially periodic potential (which mimics the ponderomotive potential of the probe pulse) to diffract the electron wave packet. The simulated momentum distributions show fringes (see Figs. 2-3) that agree very well with the experimental observation and its time-domain behavior.

The time-resolved Kapitza-Dirac effect can be conceptually understood as an interference between the initial electron wave packet and a replica of the electron wave packet that has been shifted in momentum space by $2k_\gamma$ resulting from stimulated Compton backscattering of the photons of the probe pulse. While the initial electron wave packet evolves under field-free conditions (thus, accompanied with a phase evolution of $\varphi_e(t) = p_x^2 t/2$), the altered electrons experience the aforementioned $2k_\gamma$ recoil along the light-propagation direction when interacting with the standing wave. Accordingly, this interaction leads to a momentum shifted electron wave packet with an accompanying phase of $\varphi'_e(t) = (p_x \pm 2k_\gamma)^2 t/2 + \varphi_{SW}$, where $\varphi_{SW}$ is the relative phase of the standing wave with respect to the initial electron wave packet (see Supplementary Materials). The tiny change of energy of the electron is compensated by an equivalent change of the photon energy upon scattering, which is within the bandwidth of the short pulse. In our case, $\varphi_{SW} = \pi$. The initial and shifted wave packets interfere constructively when the phase difference is $\Delta\varphi_e = \varphi'_e - \varphi_e = 2n\pi$, where $n$ is an integer. Thus, the peaks in the final electron momentum distribution are expected to be located at $(p_x)_n \approx \frac{n\pi}{tk_\gamma} = (n + \frac{1}{2})\frac{\lambda_{SW}}{2t}$, resulting in a modulation-spacing of $\Delta p_x = \frac{\lambda_{SW}}{2t}$.

Finally, we would like to stress the following implications of our findings. The time-resolved Kapitza-Dirac effect differs fundamentally from the conventional Kapitza-Dirac effect. In momentum space, the latter generates extremely narrow (delta function-like) side bands in momentum space. As these sidebands and the original momentum distribution of the electron do not overlap an interference of these contributions is not possible. In case of the time-resolved Kapitza-Dirac effect, however, the momentum shift of $2k_\gamma$ is much smaller than the width of the initial electron wave packet, which enables an interference of the initial and scattered contribution (Figs. 2-3). The observed interference fringes do not only provide insight into the temporal dynamics of the studied light-matter interaction. We envision that the time-resolved Kapitza-Dirac effect can be used as a powerful interferometer to measure phase gradients of free electrons in general, an observable which is otherwise inaccessible using state-of-the-art electron spectroscopy. Techniques like velocity map imaging, dispersive electron analyzers and COLTRIMS reaction microscopes routinely access the amplitude of electron wave packets. Approaches to learn about the phase of photoelectrons such as RABBITT (*20*, *21*) or HASE (*22*, *23*) rely on intercepting the photoelectron's creation mechanism. In contrast, the time-resolved Kapitza-Dirac effect acts as an electron interferometer that reads out phase information long after the ionization event, leaving the ionization process undistorted. The interference fringes shown in Figs. 2-3 for positive time delays are therefore results of probing a freely propagating electron and accessing its quantum-phase properties. Possible applications of such novel pulsed Kapitza-Dirac interferometers consist, for example, of probing of phase entanglement between several electrons or between electrons and ions. Other possible future applications are probing of electrons released from surfaces or of interrogating chiral electron wave packets tracking the origin of their chirality as they emerge from chiral molecules.

**References and Notes**

1. C. Davisson, L. H. Germer, The Scattering of Electrons by a Single Crystal of Nickel. *Nature*. **119**, 558–560 (1927).

2. P. L. Kapitza, P. A. M. Dirac, The reflection of electrons from standing light waves. *Math. Proc. Cambridge Philos. Soc.* **29**, 297–300 (1933).

3. H. Batelaan, Colloquium: Illuminating the Kapitza-Dirac effect with electron matter optics. *Rev. Mod. Phys.* **79**, 929–941 (2007).

4. P. L. Gould, G. A. Ruff, D. E. Pritchard, Diffraction of atoms by light: The near-resonant Kapitza-Dirac effect. *Phys. Rev. Lett.* **56**, 827–830 (1986).

5. O. Nairz, B. Brezger, M. Arndt, A. Zeilinger, Diffraction of Complex Molecules by Structures Made of Light. *Phys. Rev. Lett.* **87**, 160401 (2001).

6. B. A. Stickler, M. Diekmann, R. Berger, D. Wang, Enantiomer Superpositions from Matter-Wave Interference of Chiral Molecules. *Phys. Rev. X*. **11**, 31056 (2021).

7. D. M. Giltner, R. W. McGowan, S. A. Lee, Atom Interferometer Based on Bragg Scattering from Standing Light Waves. *Phys. Rev. Lett.* **75**, 2638–2641 (1995).

8. X. Wu, Z. Pagel, B. S. Malek, T. H. Nguyen, F. Zi, D. S. Scheirer, H. Müller, Gravity surveys using a mobile atom interferometer. *Sci. Adv.* **5**, eaax0800 (2019).

9. M. M. Dellweg, C. Müller, Spin-Polarizing Interferometric Beam Splitter for Free Electrons. *Phys. Rev. Lett.* **118**, 70403 (2017).

10. D. L. Freimund, K. Aflatooni, H. Batelaan, Observation of the Kapitza-Dirac effect. *Nature*. **413**, 142–143 (2001).

11. D. L. Freimund, H. Batelaan, Bragg Scattering of Free Electrons Using the Kapitza-Dirac Effect. *Phys. Rev. Lett.* **89**, 283602 (2002).

12. P. H. Bucksbaum, D. W. Schumacher, M. Bashkansky, High-Intensity Kapitza-Dirac Effect. *Phys. Rev. Lett.* **61**, 1182–1185 (1988).

13. X. Li, J. Zhang, Z. Xu, P. Fu, D.-S. Guo, R. R. Freeman, Theory of the Kapitza-Dirac Diffraction Effect. *Phys. Rev. Lett.* **92**, 233603 (2004).

14. S. Eilzer, H. Zimmermann, U. Eichmann, Strong-field Kapitza-Dirac scattering of neutral atoms. *Phys. Rev. Lett.* **112**, 113001 (2014).

15. A. Hartung, S. Eckart, S. Brennecke, J. Rist, D. Trabert, K. Fehre, M. Richter, H. Sann, S. Zeller, K. Henrichs, G. Kastirke, J. Hoehl, A. Kalinin, M. S. Schöffler, T. Jahnke, L. P. H. Schmidt, M. Lein, M. Kunitski, R. Dörner, Magnetic fields alter strong-field ionization. *Nat. Phys.* **15**, 1222–1226 (2019).

16. R. Dörner, V. Mergel, O. Jagutzki, L. Spielberger, J. Ullrich, R. Moshammer, H. Schmidt-Böcking, Cold Target Recoil Ion Momentum Spectroscopy: a 'momentum microscope' to view atomic collision dynamics. *Phys. Rep.* **330**, 95–192 (2000).

17. T. Brabec, M. Y. Ivanov, P. B. Corkum, Coulomb focusing in intense field atomic processes. *Phys. Rev. A*. **54**, R2551–R2554 (1996).
5

**Acknowledgments:** K. L. acknowledges Chenxu Lu and Jinliang Jiang for figure plotting, and Peilun He and Feng He for fruitful discussion.

**Funding:** S. E. acknowledges funding of the DFG through Priority Programme SPP 1840 QUTIF. The experimental work was supported by the DFG (German Research Foundation).

**Author contributions:** K.L., S.E., A.H., S.J., Q.J., L.Ph.H.S., M.S.S., T.J., M.K. and R.D. contributed to the experiment. K.L., S.E., M.K. and R.D. did the data analysis. M.K. and H.L. performed simulation. All authors contributed to the manuscript.

**Competing interests:** Authors declare that they have no competing interests.

**Data and materials availability:** All data are available in the main text or the supplementary materials.

**Supplementary Materials**

Materials and Methods

Supplementary Text

Figs. S1 to S3

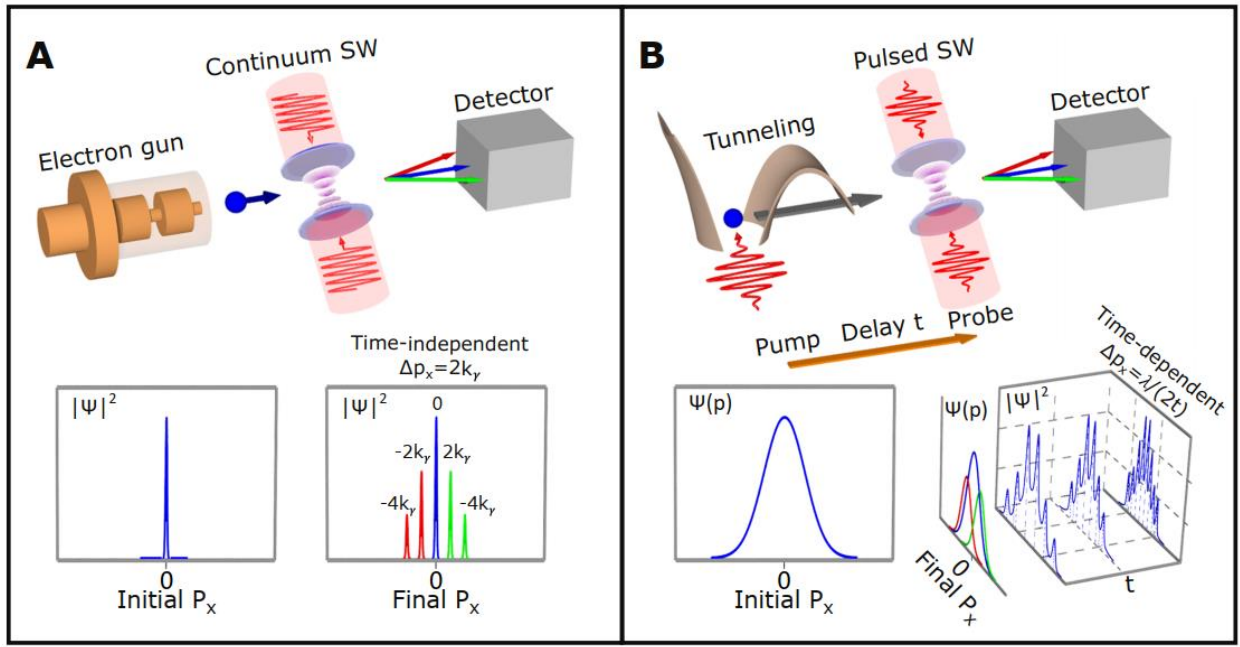

**Fig. 1. Momentum space perspective of the conventional and the time-resolved Kapitza-Dirac effects.** (**A**) The conventional Kapitza-Dirac effect describes the diffraction of a plane electron wave that can be represented as a delta function in momentum space. Thus, the width of the momentum distribution is much narrower than two-photon momenta $2k_\gamma$. In a standing light wave stimulated Compton backscattering of photons shifts this narrow momentum distribution by $\pm 2nk_\gamma$, where $n$ is an integer. The resulting regularly spaced peak structure in momentum space is time-independent. (**B**) In contrast, for the time-resolved Kapitza-Dirac effect the stimulated Compton backscattering transfers the momentum to an electron wave packet, that has a very broad momentum distribution. The Compton backscattering occurs during a short time interval (femtoseconds) and at a variable time delay after the birth of the wave packet. The interference between the original (blue Gaussian) and the momentum shifted (red and green Gaussians) wave packets leads to time-dependent fringes in momentum space.

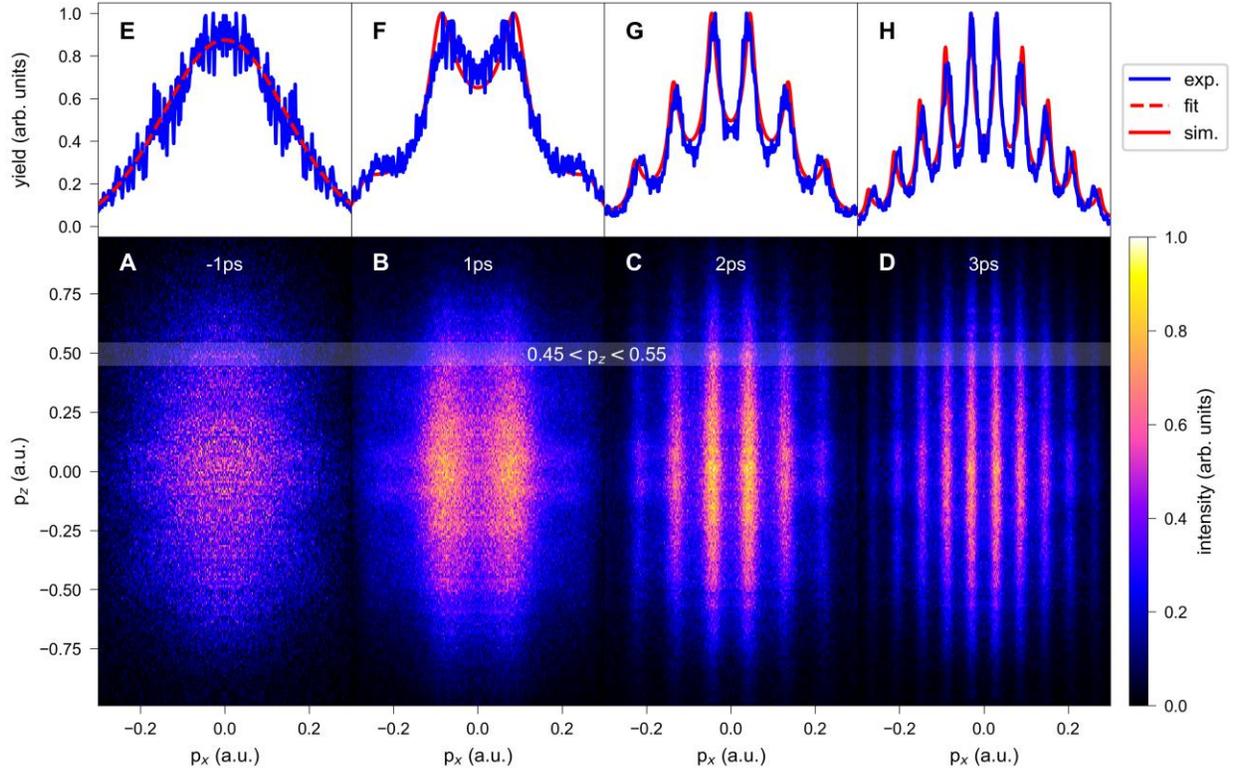

**Fig. 2. Observation of the time-resolved Kapitza-Dirac effect**. (**A-D**) Experimentally measured momentum distributions of the electron that is released upon strong-field ionization of xenon atoms in a pulsed standing light wave (800 nm, 60 fs, $1.0\times10^{14}$ W/cm$^2$). The ionizing pulse is followed by a weak standing light wave pulse ($0.4\times10^{14}$ W/cm$^2$) that acts as a probe pulse. The probe pulse is attenuated such that it cannot lead to the liberation of bound electrons and thus only serves as a diffraction grating. Vertical axis: electron momentum along the polarization direction ($p_z$); Horizontal axis: electron momentum along the light-propagation direction ($p_x$). Panels (A-D) show the data taken for different time delays of -1, 1, 2, and 3 ps, respectively. (**E-H**) One-dimensional momentum distribution along the light-propagation axis by selecting the range of $p_z$ between (0.45, 0.55) atomic units (shaded area in panels (A-D)). The dashed curve in (E) is obtained by fitting a Gaussian distribution, while the solid curves in (F-H) show results from numerically propagating one dimensional electron wave packet followed by interaction with standing wave at different time delays.

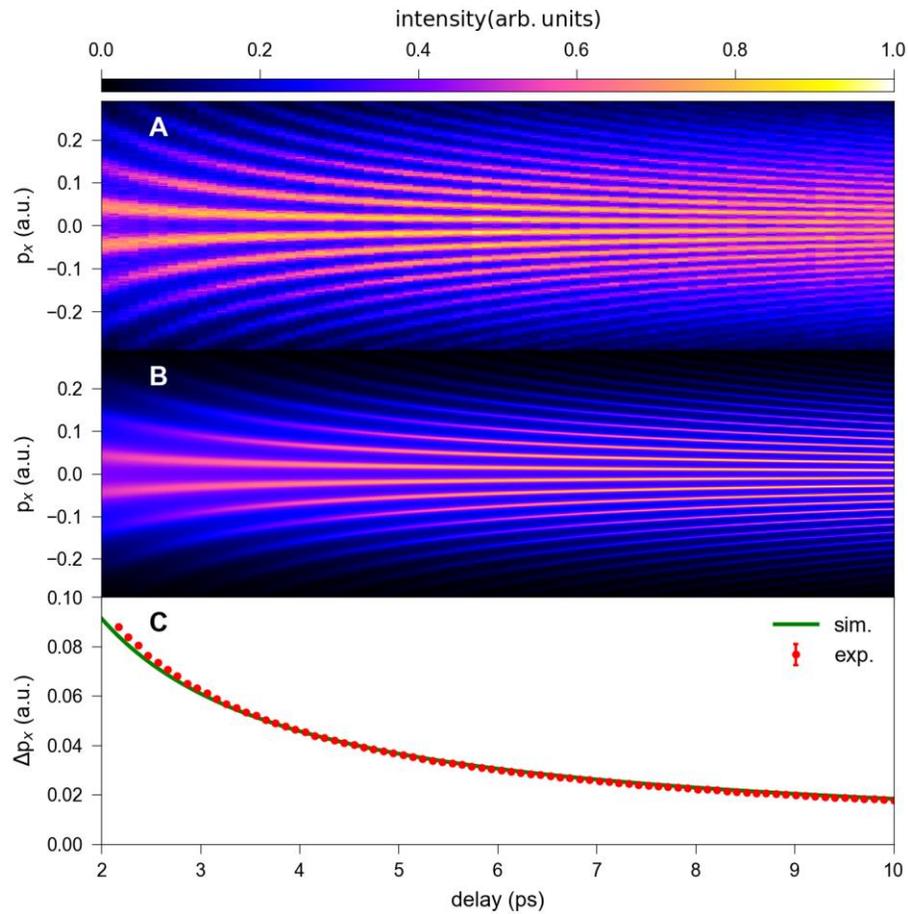

**Fig. 3. Diffraction pattern as a function of time delay**. (**A**) Measured and (**B**) simulated electron momentum distribution along $p_x$ for light pulses as in Fig. 2 as a function of the time delay between ionization (by the pump pulse) and probing by the nonionizing standing light wave (probe pulse). The data are integrated over $|p_y| > 0.1$ a.u. (**C**) Fitted diffraction spacing from (A) as a function of the time delay. The solid curve is obtained from our simulations in full analogy to the experiment.